\documentclass[11pt]{article}

\usepackage{epsfig}
\usepackage{graphicx}

\begin{document}
\title{Electronic properties of graphene and graphene nanoribbons with ``pseudo-Rashba'' spin-orbit coupling}
\author{Tobias Stauber$^{1}$ and John Schliemann$^{2}$\\\\
$^{1}${\it Departamento de Física and Centro de Física, 
Universidade do Minho,}\\
{\it P-4710-057 Braga, Portugal}\\
$^{2}${\it Institute for Theoretical Physics, University of Regensburg,}\\ 
{\it D-93040 Regensburg, Germany}}
\date{\today}
\maketitle
\abstract{We discuss the electronic properties of graphene and graphene nanoribbons including ``pseudo-Rashba'' spin-orbit coupling. After summarizing the bulk properties, we first analyze the scattering behavior close to an infinite mass and zigzag boundary. For low energies, we observe strong deviations from the usual spin-conserving behavior at high energies such as reflection acting as spin polarizer or switch. This results in a spin polarization along the direction of the boundary due to the appearance of evanescent modes in the case of non-equilibrium or when there is no coherence between the two one-particle branches. We then discuss the spin and density distribution of graphene nanoribbons.}\\
\\
{Pacs: 71.70.Ej,73.61.Wp,72.25.Rb}
\section{Introduction}

Graphene, the single-layer allotrope of carbon, is undoubtedly one of the most
active fields in today\rq s both experimental
and theoretical condensed matter physics 
\cite{Novoselov04,Geim07,CastroNeto07}. Among an entire plethora of
phenomena and proposals, the issue of spin-orbit coupling has generated 
particular interest \cite{Kane05,Min06,Huertas06,Yao07,Boettger07,Gmitra09}.
A detailed understanding of spin-orbit interaction in graphene is crucial
for the interpretation of ongoing experiments on spin transport
performed by various groups
\cite{Hill06,Tombros07,Nishioka07,Cho07,Ohishi07,Josza08,Wang08,Goto08,Han09}.
Other issues include various device proposals \cite{Trauzettel07,Ezawa09}
and theoretical predictions \cite{Brey07,Ding08,Onari08} related to
spins and spin-orbit coupling in graphene.

In the present paper we investigate a single layer of graphene in the presence
of spin-orbit interaction of the  ``pseudo-Rashba'' type, coupling the
sublattice or pseudo spin to the physical electron spin
\cite{Kane05,Min06,Huertas06,Yao07,Boettger07,Gmitra09,Rashba09}. Our interest is based on the fact that for graphene on Ni with intercalation of Au a 100-fold enhancement of the ``pseudo-Rashba'' spin-orbit coupling has been reported \cite{Varykhalov08}. Furthermore, impurities which induce a sp3-distortion will lead to a  ``pseudo-Rashba'' spin-orbit coupling with a value comparable to the one found in diamond and other zinc-blende semiconductors \cite{Neto09}. The latter result indicates that the ``pseudo-Rashba'' spin-orbit coupling can be controlled via the impurity coverage.

In this paper, we will concentrate on the scattering behavior of spin
densities near boundaries created either by an infinite mass or a
zigzag edge.  Our presentation is organized as follows: In section
\ref{general} we introduce the basic Hamiltonian and discuss its
general bulk solution in the absence of a mass term; the technically more
complicated case of a nonzero mass is deferred to the appendices. In
the following section \ref{dephasing} we investigate in detail the
scattering properties and spin dephasing at hard boundaries for
various types of incoming spinors and energy ranges. This discussion
is extended in section \ref{polarization} to averaged spin
polarizations obtained from continuous distributions of incoming
directions. In section \ref{sec:ribbon}, we analyze the spin and
density distribution of graphene nanoribbons. We close with a summary
in section \ref{summary}. Throughout this manuscript, we use
parameters of Ref. \cite{Varykhalov08}.

\section{Dirac fermions with ``pseudo-Rashba'' spin-orbit coupling}
\label{general}

The single-particle Hamiltonian of monolayer graphene with 
 ``pseudo-Rashba'' spin-orbit interaction can be formulated as
\cite{Kane05,Min06,Huertas06,Rashba09}
\begin{equation}
{\cal H}=v_F\vec p\cdot\vec\tau+\lambda\left(\vec\tau\times\vec\sigma\right)
\cdot\vec e_{z}\,,
\end{equation}
where, among standard notation, $\lambda$ is the spin-orbit coupling parameter,
and the Pauli matrices $\vec\tau$, $\vec\sigma$ describe the 
sublattice and the electron spin degree of freedom, respectively. For a given
wave vector $\vec k$ this Hamiltonian reads explicitly:
\begin{equation}
{\cal H}(\vec k)=
\left(
\begin{array}{cccc}
 0 & 0 & \hbar v_F(k_{x}-ik_{y}) & 0 \\
 0 & 0 & 2i\lambda & \hbar v_F(k_{x}-ik_{y}) \\
\hbar v_F(k_{x}+ik_{y}) & -2i\lambda & 0 & 0\\
 0 & \hbar v_F(k_{x}+ik_{y}) & 0 & 0 
\end{array}
\right)
\end{equation}
From experience with the ``classic'' Dirac equation of relativistic
quantum mechanics, it is occasionally of use not to study just a given
Hamiltonian but also its square. Here we find
\begin{equation}
{\cal H}^{2}(\vec k)=
\left(
\begin{array}{cccc}
(\hbar v_Fk)^{2} & -2i\lambda\hbar v_F(k_{x}-ik_{y}) & 0 & 0 \\
2i\lambda\hbar v_F(k_{x}+ik_{y}) & (\hbar v_Fk)^{2}+4\lambda^{2} & 0 & 0 \\
 0 & 0 & (\hbar v_Fk)^{2}+4\lambda^{2} & -2i\lambda\hbar v_F(k_{x}-ik_{y}) \\
 0 & 0 & 2i\lambda\hbar v_F(k_{x}+ik_{y}) & (\hbar v_Fk)^{2}\\
\end{array}
\right)
\end{equation}
This matrix is block-diagonal with eigenvalues
\begin{equation}
(\varepsilon^{2})_{1,2}=(\hbar v_Fk)^{2}+2\lambda^{2}
\pm 2|\lambda|\sqrt{(\hbar v_Fk)^{2}+\lambda^{2}}
\end{equation}
where the positive sign corresponds to the eigenvectors
\begin{equation}
\label{alpha_beta_1}
|\alpha_{1}\rangle=\left(
\begin{array}{c}
\sin(\vartheta/2)\\
\cos(\vartheta/2)e^{i\eta}\\
0\\
0
\end{array}
\right)\qquad,\qquad|\beta_{1}\rangle=\left(
\begin{array}{c}
0\\
0\\
\cos(\vartheta/2)\\
\sin(\vartheta/2)e^{i\eta}
\end{array}
\right)\,,
\end{equation}
while for the negative sign we have
\begin{equation}
\label{alpha_beta_2}
|\alpha_{2}\rangle=\left(
\begin{array}{c}
\cos(\vartheta/2)\\
-\sin(\vartheta/2)e^{i\eta}\\
0\\
0
\end{array}
\right)\qquad,\qquad|\beta_{2}\rangle=\left(
\begin{array}{c}
0\\
0\\
-\sin(\vartheta/2)\\
\cos(\vartheta/2)e^{i\eta}
\end{array}
\right)
\end{equation}
where $\vartheta\in[0,\pi]$ and 
\begin{equation}
\cos\vartheta=\frac{|\lambda|}{\sqrt{(\hbar v_Fk)^{2}+\lambda^{2}}}\qquad,\qquad
e^{i\eta}=\frac{\lambda}{|\lambda|}\frac{i(k_{x}+ik_{y})}{k}\,.
\end{equation}
In the basis $(|\alpha_{1}\rangle,|\beta_{1}\rangle,
|\alpha_{2}\rangle,|\beta_{2}\rangle)$ the Hamiltonian reads
\begin{equation}
\tilde{\cal H}(\vec k)=
\left(
\begin{array}{cccc}
 0 & q_{+}^{\ast} & 0 & 0 \\
 q_{+} & 0 & 0 & 0 \\
 0 & 0 & 0 &  q_{-} \\
 0 & 0 & q_{-}^{\ast} & 0 
\end{array}
\right)
\end{equation}
with
\begin{equation}
q_{\pm}=\pm\hbar v_F(k_{x}\pm ik_{y})f_{\pm}(|\lambda|/ \hbar v_Fk)
\end{equation}
and 
\begin{equation}
f_{\pm}(x)=\sqrt{1+x^{2}}\pm x\,.
\end{equation}
Now it is straightforward to obtain the full eigensystem:
We find a gaped pair of eigenvalues
\begin{equation}
\varepsilon_{1,\pm}=\pm\left(\sqrt{(\hbar v_Fk)^{2}+\lambda^{2}}+|\lambda|\right)
\end{equation}
with eigenspinors (type I)
\begin{equation}
|\chi_{1,\pm}(\vec k)\rangle=
\frac{1}{\sqrt{2}}\left(
\begin{array}{c}
\sin(\vartheta/2)\\
\cos(\vartheta/2)e^{i\eta}\\
\pm\cos(\vartheta/2)e^{i\psi}\\
\pm\sin(\vartheta/2)e^{i\eta}e^{i\psi}
\end{array}
\right)
\end{equation}
and 
\begin{equation}
e^{i\psi}=\frac{k_{x}+ik_{y}}{k}\,.
\end{equation}
With $g_V=2$ being the valley degeneracy, 
the corresponding density of states reads 
\begin{equation}
\label{rhoI}
\rho_1(\varepsilon)=\frac{g_V}{2\pi(\hbar v_F)^2}\left(|\varepsilon|-|\lambda|\right)\theta\left(\varepsilon^2-(2\lambda)^2\right)\;.
\end{equation}
The other pair of dispersion branches does not exhibit a gap,
\begin{equation}
\varepsilon_{2,\pm}=\pm\left|\sqrt{(\hbar v_Fk)^{2}+\lambda^{2}}-|\lambda|\right|
\end{equation}
and has eigenspinors (type II)
\begin{equation}
|\chi_{2,\pm}(\vec k)\rangle=
\frac{1}{\sqrt{2}}\left(
\begin{array}{c}
\cos(\vartheta/2)\\
-\sin(\vartheta/2)e^{i\eta}\\
\pm\sin(\vartheta/2)e^{i\psi}\\
\mp\cos(\vartheta/2)e^{i\eta}e^{i\psi}
\end{array}
\right)\,.
\end{equation}
The corresponding density of states reads
\begin{equation}
\label{rhoII}
\rho_2(\varepsilon)=\frac{g_V}{2\pi(\hbar v_F)^2}\left(|\varepsilon|+|\lambda|\right)\;.
\end{equation}
Let us now consider expectation values within the eigenstates with
wave functions
\begin{equation}
\langle\vec r|\vec k,\mu,\pm\rangle=\frac{e^{i\vec k\vec r}}{\sqrt{\cal A}}
|\chi_{\mu,\pm}\rangle\,,
\end{equation}
$\mu\in\{1,2\}$, and $\cal A$ being the area of the system.
Here we find
\begin{equation}
\langle\vec k,1,\pm|\vec\tau|\vec k,1,\pm\rangle
=\langle\vec k,2,\pm|\vec\tau|\vec k,2,\pm\rangle
=\pm\frac{\lambda}{|\lambda|}\left(
\begin{array}{c}
\sin\vartheta\cos\varphi\\
\sin\vartheta\sin\varphi\\
0
\end{array}
\right)\,,
\end{equation}
and
\begin{equation}
\langle\vec k,1,\pm|\vec\sigma|\vec k,1,\pm\rangle
=-\langle\vec k,2,\pm|\vec\sigma|\vec k,2,\pm\rangle
=\frac{\lambda}{|\lambda|}\left(
\begin{array}{c}
-\sin\vartheta\sin\varphi\\
\sin\vartheta\cos\varphi\\
0
\end{array}
\right)\,.
\end{equation}
Here, $\varphi$ is the usual azimuthal angle of the wave 
vector, $\vec k=k(\cos\varphi,\sin\varphi)$. Note that 
\begin{equation}
\langle\vec\tau\rangle\cdot\langle\vec\sigma\rangle
=\vec k\cdot\langle\vec\sigma\rangle=0\,,
\end{equation}
as usual for Rashba spin-orbit coupling, and
\begin{equation}
|\langle\vec\tau\rangle|=|\langle\vec\sigma\rangle|=\sin\vartheta\,,
\end{equation}
where for $\sin\vartheta<1$ sublattice and electron spin degree of 
freedom are entangled which each other.

\section{Spin dephasing due to reflection on a hard wall}
\label{dephasing}
\begin{figure}[t]
\begin{center}
\includegraphics[angle=0,width=0.5\linewidth]{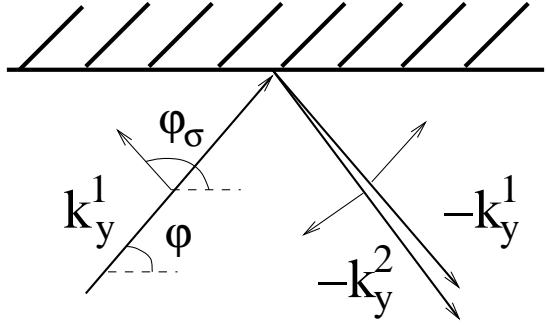}
\caption{A plane wave of type I with spin perpendicular to the momentum $\vec k=(k_x,k_y^1)$ ($\varphi=\arctan(k_y^1/k_x)$, $\varphi_\sigma=\varphi+\pi/2$) is reflected at the boundary into a plane wave with ${\vec k}^\prime=(k_x,-k_y^1)$ and  ${\vec k}^{\prime\prime}=(k_x,-k_y^2)$ with perpendicular spin, but anti-parallel with respect to each other (see Eq. (\ref{Def:ky}) for the definition of $k_y^{1/2}$).}
  \label{fig:scatter}
\end{center}
\end{figure}

In this section, we will study the scattering behavior from a hard wall which will lead to spin dephasing as depicted in Fig. \ref{fig:scatter}. For that, a general plane wave with fixed momentum $k_x$ and energy $E\geq2|\lambda|$ is written as
\begin{eqnarray}
\psi_{E,k_x}(x,y)&=&\mathcal{N}_{\vec k}e^{ik_xx}\Big[A_1e^{ik_y^1y}|\chi_{1,+}(k_x,k_y^1)\rangle +A_2e^{ik_y^2y}|\chi_{2,+}(k_x,k_y^2)\rangle\nonumber\\
\label{wavefunction}
    &+&R_1e^{-ik_y^1y}|\chi_{1,+}(k_x,-k_y^1)\rangle +R_2e^{-ik_y^2y}|\chi_{2,+}(k_x,-k_y^2)\rangle\Big]\;,
\end{eqnarray}
with
\begin{equation}
\label{Def:ky}
(\hbar v_Fk_y^\mu)^2=(E+(-1)^\mu|\lambda|)^2-\lambda^2-(\hbar v_Fk_x)^2
\end{equation}
$\mu\in\{1,2\}$ and the normalization constant $\mathcal{N}_{\vec k}$. For energies $E<2|\lambda|$, some modification in Eq. (\ref{wavefunction}) have to be made which shall be discussed in more detail below.

In the following, we will discuss the reflection at a hard wall at $y=0$ for 
the two types of plane waves, i.e., we will first set $A_1=1$, $A_2=0$ (type I) and then $A_1=0$, $A_2=1$ (type II). The discussion is based on the reflected 
spin direction which shall be denoted by $\varphi_\sigma'$. It is obtained 
from the expectation value of the spin-density operator at the boundary
$\vec\rho=\vec\sigma\delta(\hat{\vec r})$, 
$\langle\vec\rho\rangle\equiv\langle \psi_{E,k_x}|\vec\rho|\psi_{E,k_x}\rangle$ via 
\begin{equation}
\label{ReflectedAngle}
\varphi_\sigma'=\arctan(\langle\rho_y\rangle/\langle\rho_x\rangle)
+\pi\theta(-\langle\rho_x\rangle)\;.
\end{equation}
Due to translational invariance in $x$-direction, $\langle\vec\rho\rangle$
will only depend on the $y$-coordinate. For the following discussion, we will also discuss the at $\vec r=0$ normalized expectation value $\langle\vec \sigma\rangle=\langle\vec\rho\rangle/\langle n\rangle$ with $\langle n\rangle\equiv\langle \psi_{E,k_x}|\delta(\hat{\vec r})|\psi_{E,k_x}\rangle$. This shall not be confused with the bulk expectation of $\vec \sigma$ as it appears in the Hamiltonian. 

We will distinguish the
two different cases of the half-plane $y\geq 0$ (scattering from the lower or 
bottom boundary) and $y\leq 0$ (scattering from the upper or top boundary). We shall further assume a plane wave with $k_x>0$ moving in positive $x$-direction. The results for $k_x<0$ are then obtained by changing bottom to top boundary and vice versa. The results for the $K'$-point can also be deduced from the following discussion (see appendix A).
The sign of $\lambda$ determines the sign of the expectation value of $\vec \tau$ and $\vec \sigma$. In the following, we set $\lambda=|\lambda|$, but in some of the following expression we explicitly use $|\lambda|$ for sake of clarity.  

We will discuss two different types of confinement. First, we use the fact that Dirac fermions can be confined by an infinite mass boundary, first discussed by Berry and Mondragon \cite{Berry87}. We then also study the reflection from a zigzag boundary first addressed in Ref. \cite{Huertas08}. 

\subsection{Infinite mass boundary}

With $\psi_{E,k_x}=(\psi_1,\psi_2,\psi_3,\psi_4)^T$, the boundary conditions at the infinite mass boundary read (see appendix B and C)
\begin{equation}
\label{BoundaryCond}
\frac{\psi_1}{\psi_3}\Bigg|_{\rm bottom}=\frac{\psi_2}{\psi_4}\Bigg|_{\rm bottom}=1\quad,\quad 
\frac{\psi_1}{\psi_3}\Bigg|_{\rm top}=\frac{\psi_2}{\psi_4}\Bigg|_{\rm top}=-1\quad.
\end{equation}
Note that there are different boundary conditions depending on whether one approaches the boundary from below or above. 

\subsubsection{Scattering behavior for plane waves of type I}

We first consider a plane wave scattered at $y=0$ with $A_1=1$ and $A_2=0$. The boundary conditions yield the following expressions for $R_1$, $R_2$:
\begin{eqnarray}
\label{A1R1}
R_1&=&\mp z_1^2\frac{(z_1c_1\pm s_1)(z_2s_2\pm c_2)+(z_1s_1\pm c_1)(z_2c_2\pm s_2)z_1z_2}{(z_1s_1\pm c_1)(z_2s_2\pm c_2)z_1+(z_1c_1\pm s_1)(z_2c_2\pm s_2)z_2}\\
\label{A1R2}
R_2&=&\mp z_2^2\frac{(z_1c_1\pm s_1)^2-(z_1s_1\pm c_1)^2z_1^2}{(z_1s_1\pm c_1)(z_2s_2\pm c_2)z_1+(z_1c_1\pm s_1)(z_2c_2\pm s_2)z_2}  
\end{eqnarray}
Above, we introduced the abbreviations $c_\mu=\cos(\vartheta_\mu/2)$, $s_\mu=\sin(\vartheta_\mu/2)$, and $z_\mu=(k_x+ik_y^\mu)/\sqrt{k_x^2+(k_y^\mu)^2}$, $\mu\in\{1,2\}$. The upper (lower) sign holds if the electron is scattered from the upper (lower) boundary.

\begin{figure}[t]
\begin{center}
\includegraphics[angle=0,width=0.5\linewidth]{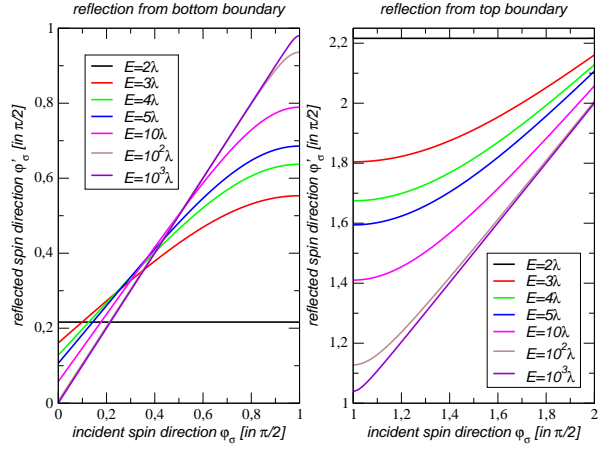}
\caption{The reflected versus the incident spin direction at $y=0$ for an incident plane wave with $A_2=0$ (type I) for various energies $E$. We use $\hbar v_F=5.6$eV\AA${}$ and $\lambda=6$meV. Left: Reflection from the lower boundary. Right: Reflection from the upper boundary.}
  \label{fig:SpinReflection1}
\end{center}
\end{figure}

Let us first discuss the scattering behavior from the lower boundary. For $k_x=k\cos\varphi$, the incident spin direction is given by $\varphi_\sigma=\pi/2-|\varphi|$. On the left hand side of Fig. \ref{fig:SpinReflection1}, the reflected spin direction $\varphi_\sigma'$ of Eq. (\ref{ReflectedAngle}) is plotted against the incident spin direction $\varphi_\sigma$.

At large energies with $\epsilon=\lambda/(\hbar v_Fk)\ll1$ and $\epsilon\ll\sin^2\varphi$, we have $R_1=(1-\epsilon)\cos\varphi$ and $R_2=i\sin\varphi-2\epsilon\cos\varphi$ and the spin polarization is approximately conserved. The expansion of Eq. (\ref{ReflectedAngle}) yields 
\begin{equation}
\varphi_\sigma'=\varphi_\sigma+\epsilon\frac{\cos\varphi_\sigma}{1+\sin\varphi_\sigma}.
\end{equation}

For energies close to the band gap energy of the type I-spinors, $E\rightarrow2\lambda$, scattering from the boundary acts as a spin polarizer since $\varphi_\sigma'\rightarrow\varphi_0=\arctan(1/(2\sqrt{2}))\approx19.5^\circ$ for all incoming spin directions $\varphi_\sigma$. This angle corresponds to $\langle\sigma_y\rangle=1/3$. For $E=\lambda(2+\epsilon^2)$ with $\epsilon\ll1$, we get
\begin{equation}
\label{Asym2L}
\varphi_\sigma'=\arctan\left(\frac{1}{2\sqrt{2}}\right)+\frac{2}{3} \epsilon \cos\varphi+\frac{\sqrt{2}}{72}\epsilon^2\left(\cos(2\varphi)-5\right)\;. 
\end{equation}
This is a surprising result since $R_1\rightarrow-1$ and incoming and reflected wave seem to compensate. But even though $R_2\rightarrow-\sqrt{6}\epsilon e^{-i\varphi}\sin\varphi$ tends to zero, its admixture has a dominating effect.

For the upper boundary, we obtain the expansion
\begin{equation}
\varphi_\sigma'=\pi+\arctan\left(\frac{1}{2\sqrt{2}}\right)-\frac{2}{3} \epsilon \cos\varphi+\frac{\sqrt{2}}{72}\epsilon^2\left(\cos(2\varphi)-5\right)\;. 
\end{equation}
Note that the different sign compared to Eq. (\ref{Asym2L}) results in a different asymptotic behavior for large energies since $\varphi_\sigma'(E=2\lambda)$ is larger than the maximal incident spin direction $\varphi_\sigma=\pi$. This different behavior is illustrated on the right hand side of Fig. \ref{fig:SpinReflection1}. 

\subsubsection{Scattering behavior for plane waves of type II with $E\geq2\lambda$}

For a plane wave scattered at $y=0$ with $A_1=0$ and $A_2=1$ with energy $E\geq2\lambda$, the boundary conditions yield the following expressions for $R_1$, $R_2$:
\begin{eqnarray}
  \label{A2R1}
R_1&=&\mp z_1^2\frac{(z_2s_2\pm c_2)^2-(z_2c_2\pm s_2)^2z_2^2}{(z_1s_1\pm c_1)(z_2s_2\pm c_2)z_1+(z_1c_1\pm s_1)(z_2c_2\pm s_2)z_2}\\  
\label{A2R2}
R_2&=&\mp z_2^2\frac{(z_1c_1\pm s_1)(z_2s_2\pm c_2)+(z_1s_1\pm c_1)(z_2c_2\pm s_2)z_1z_2}{(z_1s_1\pm c_1)(z_2s_2\pm c_2)z_1+(z_1c_1\pm s_1)(z_2c_2\pm s_2)z_2}
\end{eqnarray}

For $(E-2|\lambda|)/(E+2|\lambda|)>(\cos\varphi)^2$, the abbreviations are the same as in Eqs. (\ref{A1R1}) and (\ref{A1R2}). For $(E-2|\lambda|)/(E+2|\lambda|)<(\cos\varphi)^2$, the reflected momentum $k_y^1=\pm iq$ is imaginary with
\begin{equation}
\label{Def:q}
\hbar v_Fq=\sqrt{-(E-|\lambda|)^2+\lambda^2+(\hbar v_Fk_x)^2}\;.
\end{equation}
The sign is determined to yield an exponential decay in the reflected region. In Eqs. (\ref{A2R1}) and (\ref{A2R2}), $z_1$ is thus replaced by $z_1\rightarrow(k_x\mp q)/\sqrt{k_x^2-q^2}$, where the upper (lower) sign holds for reflections from the upper (lower) boundary, and $s_1$ by $s_{1}\rightarrow i\sqrt{(\cos\vartheta_1-1)/2}$.

\begin{figure}[t]
\begin{center}
\includegraphics[angle=0,width=0.5\linewidth]{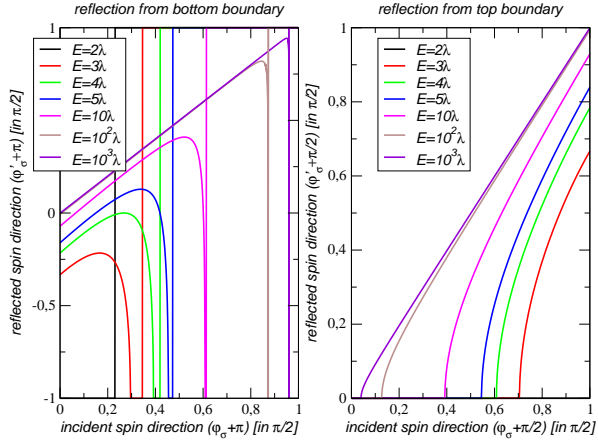}
\caption{The reflected versus the incident spin direction (rotated by $\pi$) at $y=0$ and $A_1=0$ for various energies $E\geq2\lambda$. We use $\hbar v_F=5.6$eV\AA${}$ and $\lambda=6$meV. Left: Reflection from the lower boundary. Right: Reflection from the upper boundary.}
  \label{fig:SpinReflection2}
\end{center}
\end{figure}

Let us first discuss the scattering behavior from the lower boundary. On the left hand side of Fig. \ref{fig:SpinReflection2}, the reflected spin direction is plotted against the incident spin direction rotated by $\pi$. For large energies and normal incident $\varphi\approx\pi/2$, we again obtain $\varphi_\sigma'=\varphi_\sigma$. But for nearly parallel incident such that $(E-2|\lambda|)/(E+2|\lambda|)<(\cos\varphi)^2$, we obtain $\varphi_\sigma'=\pm\pi/2$. For energies close to the band-gap $E\rightarrow2\lambda$, all reflected modes of type I are evanescent and scattering from the wall acts as a switch which leads to either $\varphi_\sigma'=\pi/2$ or $\varphi_\sigma'=-\pi/2$.

\begin{figure}[t]
\begin{center}
\includegraphics[angle=0,width=0.5\linewidth]{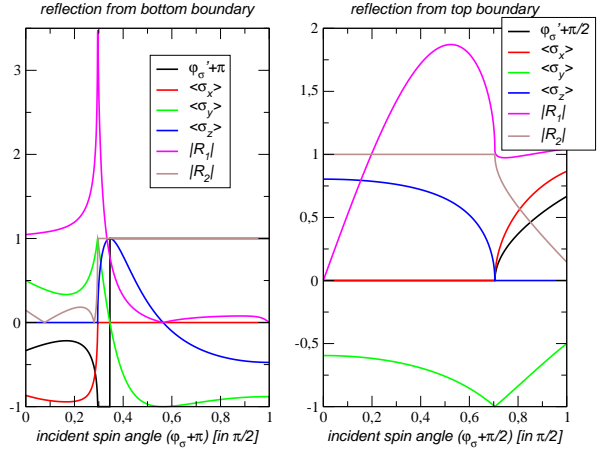}
\caption{The reflected spin angle (rotated by $\pi$), the expectation values $\langle\sigma_i\rangle$ ($i=x,y,z$) and the absolute value of the reflection amplitudes $|R_1|$ and $|R_2|$ versus the incident spin direction at $y=0$ and $A_2=0$ for energies $E=3\lambda$. We use $\hbar v_F=5.6$eV\AA${}$ and $\lambda=6$meV. Left: Reflection from the lower boundary. Right: Reflection from the upper boundary.}
  \label{fig:Example3}
\end{center}
\end{figure}

Let us understand the appearance of the two extreme values of $\varphi_\sigma'=\pm\pi/2$ in the regime where $k_y^1$ is imaginary. Since $z_1$ is real and the incident and reflected wave of type $|\chi_{2,+}\rangle$ compensate, the expectation value in $x$-direction $\langle\sigma_x\rangle=0$. For the incident wave, $\langle|\sigma_y|\rangle_{\rm incident}$ is negative and for small incident angle, we thus have $\varphi_\sigma'=-\pi/2$. But if $|R_1|$ is large, the admixture of $|\chi_{1,+}\rangle$ can lead to $\varphi_\sigma'=\pi/2$. Additionally, the spin in $z$-direction $\langle\sigma_z\rangle$ assumes a non-zero value to guarantee $|\langle\vec \sigma\rangle|=1$. On the left hand side of Fig. \ref{fig:Example3}, this general behavior is shown whether the reflected spin angle (rotated by $\pi$), the expectation values $\langle\sigma_i\rangle$ ($i=x,y,z$) and the absolute value of the reflection amplitudes $|R_1|$ and $|R_2|$ is plotted versus the incident spin direction at $y=0$ at energy $E=3\lambda$.

The scattering behavior from the upper boundary is considerably simpler. There, only two regimes appear with are marked by whether $k_y^1$ is real or imaginary. This can be seen on the right hand side of Figures \ref{fig:SpinReflection2} and \ref{fig:Example3}.

\subsubsection{Scattering behavior for plane waves of type II with $E<2\lambda$}

For energies with $E<2\lambda$, one of the reflected modes becomes evanescent which leads to $\langle\sigma_x\rangle=0$. For a more detailed analysis, we have to distinguish the two cases $E>\lambda$ and $E<\lambda$.

For $\lambda<E<2\lambda$, the reflected momentum $k_y^1=\pm iq$ is imaginary with the same expression as in Eq. (\ref{Def:q}). The sign is determined to yield an exponential decay in the reflected region. With the ansatz
\begin{eqnarray}
\psi_{E,k_x}(x,y)&=&\mathcal{N}_{\vec k}e^{ik_xx}\Big[e^{ik_y^2y}|\chi_{2,+}(k_x,k_y^2)\rangle\nonumber\\
    &+&\widetilde R_1e^{-q|y|}|\chi_{1,+}(k_x,\pm iq)\rangle +R_2e^{-ik_y^2y}|\chi_{2,+}(k_x,-k_y^2)\rangle\Big]\;,
\end{eqnarray}
we obtain the same expressions for $\widetilde R_1\rightarrow R_1$ and $R_2$ as in Eqs. (\ref{A2R1}) and (\ref{A2R2}) with the replacement $c_{1}\rightarrow\sqrt{(1+\cos\vartheta_1)/2}$, $s_{1}\rightarrow i\sqrt{(\cos\vartheta_1-1)/2}$, and $z_{1}\rightarrow-i(k_x\mp q)/\sqrt{q^2-k_x^2}$, where the upper (lower) sign holds for reflections from the upper (lower) boundary. 

Let us first discuss the lower boundary. For small incident spin direction, $\langle\sigma_y\rangle>0$ and becomes zero at $\varphi_\sigma=\varphi_E<\varphi_0\approx19.5^\circ$. The reflected spin angle is thus $\varphi_\sigma'=\pi/2$ for $\varphi>\varphi_E$ and $\varphi_\sigma'=-\pi/2$ for $\varphi<\varphi_E$ and for $E\rightarrow\lambda$ we have $\varphi_{E\rightarrow\lambda}=0$. 

For the upper boundary, we have $\langle\sigma_y\rangle<0$ for all angles and energies. In both cases, we have $\langle\sigma_z\rangle\neq0$ to fulfill the sum rule $|\langle\vec \sigma\rangle|=1$.

For energies with $0<E<\lambda$, there is no reflected wave of type I, $|\chi_{1,+}\rangle$, but one of the reflected momenta of $|\chi_{2,+}\rangle$ is imaginary, $k_y^2=\pm iq$ with the same definition as in Eq. (\ref{Def:q}). With 
\begin{eqnarray}
\psi_{E,k_x}(x,y)&=&\mathcal{N}_{\vec k}e^{ik_xx}\Big[e^{ik_y^2y}|\chi_{2,+}(k_x,k_y^2)\rangle\nonumber\\
    &+&\widetilde R_2e^{-q|y|}|\chi_{2,+}(k_x,\pm iq)\rangle +R_2e^{-ik_y^2y}|\chi_{2,+}(k_x,-k_y^2)\rangle\Big]\;,
\end{eqnarray}
we have
\begin{eqnarray}
\widetilde R_2&=&\mp \tilde z_2^2\frac{(z_2s_2\pm c_2)^2-(z_2c_2\pm s_2)^2z_2^2}{(\tilde z_2 \tilde s_2\pm \tilde c_2)(z_2c_2\pm s_2)z_2-(\tilde z_2 \tilde c_2\pm \tilde s_2)(z_2s_2\pm c_2)\tilde z_2}\;,\\  
R_2&=&\mp z_2^2\frac{(\tilde z_2 \tilde s_2\pm\tilde c_2)(z_2s_2\pm c_2)-(\tilde z_2 \tilde c_2\pm \tilde s_2)(z_2c_2\pm s_2)\tilde z_2z_2}{(\tilde z_2 \tilde s_2\pm \tilde c_2)(z_2c_2\pm s_2)z_2-(\tilde z_2 \tilde c_2\pm \tilde s_2)(z_2s_2\pm c_2)\tilde z_2}\;,
\end{eqnarray}   
with $\tilde c_{2}=\sqrt{(1+\cos\vartheta_2^e)/2}$, $\tilde s_{2}=i\sqrt{(\cos\vartheta_2^e-1)/2}$, $\tilde z_{2}=-i(k_x\mp q)/\sqrt{q^2-k_x^2}$, and $\vartheta_2^e=|\lambda|/(|\lambda|-E)$. In the above equations, the upper (lower) sign holds for reflections from the upper (lower) boundary.

We obtain $\langle\sigma_y\rangle=-1$ for the upper and $\langle\sigma_y\rangle=1$ for the lower boundary, respectively which is independent of the incident direction nor of the energy.

\begin{figure}[t]
\begin{center}
\includegraphics[angle=0,width=0.5\linewidth]{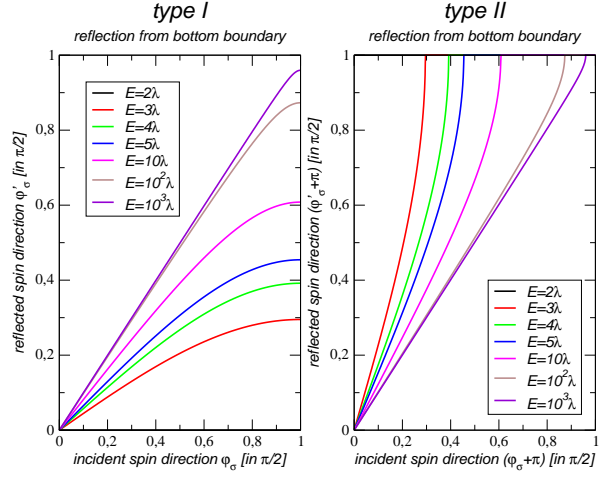}
\caption{The reflected versus the incident spin direction at $y=0$ with $A_2=0$ (left hand side) and $A_1=0$ (rotated by $\pi$) (right hand side) for various energies $E\geq2\lambda$ in the case of a zigzag boundary. We use $\hbar v_F=5.6$eV\AA${}$ and $\lambda=6$meV.}
  \label{fig:SpinReflectionZigZag}
\end{center}
\end{figure}
\subsection{Zigzag boundary}

Graphene can be terminated by a zigzag boundary which exposes only one sublattice to the boundary. 
With $\psi_{E,k_x}=(\psi_1,\psi_2,\psi_3,\psi_4)^T$, the boundary conditions at a zigzag boundary thus read
\begin{equation}
\label{BoundaryCondzigzag}
\psi_1=\psi_2=0 \quad{\rm (for}\;{\rm bottom}\;{\rm boundary)}\quad,\quad 
\psi_3=\psi_4=0 \quad{\rm (for}\;{\rm top}\;{\rm boundary)}\quad.
\end{equation}
Here, we assumed that the bottom boundary is terminated by sublattice $A$ and the top boundary by sublattice $B$.

For a general plane wave Eq. (\ref{wavefunction}) scattered at $y=0$ with energy $E\geq2\lambda$, the boundary conditions for the bottom boundary (sublattice $A$) Eq. (\ref{BoundaryCondzigzag}) yield the following expressions for $R_1$, $R_2$:
\begin{eqnarray}
  \label{R1zigzagA}
R_1&=&-z_1^2\frac{A_1(s_1s_2+c_1c_2z_1z_2)+A_2(s_2c_2-s_2c_2z_2^2)}{s_1s_2z_1^2+c_1c_2z_1z_2}\\  
\label{R2zigzagA}
R_2&=&-z_2^2\frac{A_1(s_1c_1-s_1c_1z_1^2)+A_2(c_1c_2+s_1s_2z_1z_2)}{c_1c_2z_2^2+s_1s_2z_1z_2}
\end{eqnarray}
The boundary conditions for the upper boundary (sublattice $B$) yield the following expressions for $R_1$, $R_2$:
\begin{eqnarray}
  \label{R1zigzagB}
R_1&=&-z_1^2\frac{A_1z_1(c_1c_2+s_1s_2z_1z_2)+A_2z_2(s_2c_2-s_2c_2z_2^2)}{c_1c_2z_1+s_1s_2z_2}\\  
\label{R2zigzagB}
R_2&=&-z_2^2\frac{A_1z_1(s_1c_1-s_1c_1z_1^2)+A_2z_2(s_1s_2+c_1c_2z_1z_2)}{c_1c_2z_1+s_1s_2z_2}
\end{eqnarray}
The abbreviations are the same as for the infinite mass boundary. Since the reflected angle is symmetric around normal incident, we will only discuss the reflection from the bottom boundary for $k_x>0$.

In Fig. \ref{fig:SpinReflectionZigZag}, the reflected versus the incident spin direction at $y=0$ is shown for the two types of incident plane waves. 
As in the case of the infinite mass boundary, $\langle\sigma_x\rangle=0$ for incident plane waves of type II with $\cos^2\varphi>(E-2|\lambda|)/(E+2|\lambda|)$. But in contrary to the infinity mass boundary, the spin-polarization in out-of-plane direction assumes a non-zero value even when the reflected wave of type I is extended. For this case, i.e., $k_y^1\in\mathbf{R}$, we obtain
\begin{equation}
\langle\sigma_z\rangle^I=-\frac{|\lambda|}{E+|\lambda|}\quad,\quad\langle\sigma_z\rangle^{II}=\frac{|\lambda|}{E-|\lambda|}\quad.
\end{equation}
The $K'$-point yields the opposite sign such that there is no net-polarization in $z$-direction. For energies $E<2\lambda$ a similar discussion as in the case of infinite mass boundary applies.

\section{Spin polarization close to the boundary}
\label{polarization}
\begin{figure}[t]
\begin{center}
\includegraphics[angle=0,width=0.5\linewidth]{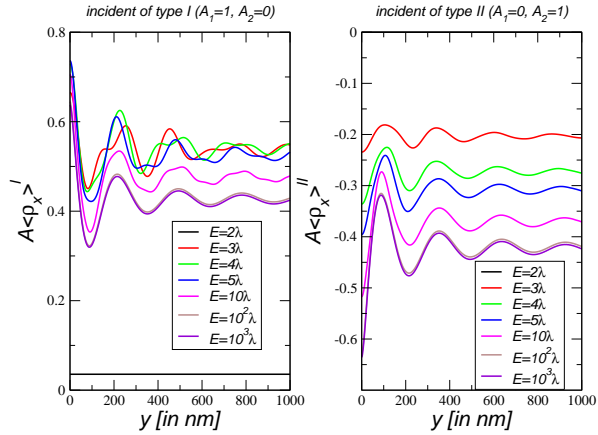}
\caption{Spin polarization in $x$-direction as function of $y$ for various energies $E\geq2\lambda$ with an infinite mass boundary. We use $\hbar v_F=5.6$eV\AA${}$ and $\lambda=6$meV. Left: Incident plane wave of type I. Right: Incident plane wave of type II.}
  \label{fig:YDependenceAve}
\end{center}
\end{figure}

So far we have only discussed the polarization properties at the boundary $y=0$. For finite $y$, we expect an oscillatory behavior of the reflected spin polarization. For $E\rightarrow2\lambda$ and plane wave scattering of type I, $k_y^1\rightarrow0$ and the period will thus be solely determined by $k_y^2\rightarrow\sqrt{2}(2\lambda/\hbar v_F)$. This oscillatory behavior is again independent of the incident spin polarization and results in a striped phase for the reflected spin-polarization. For $E>2\lambda$, two periods related to $k_y^{1/2}$ contribute and a more complicated pattern emerges which also depends on the incident spin polarization and whether one deals with a reflection from the top or from the bottom. This hints to the fact that a Dirac particle in a box shows quasi-chaotic behavior \cite{Peres09}.  

\begin{figure}[t]
\begin{center}
\includegraphics[angle=0,width=0.5\linewidth]{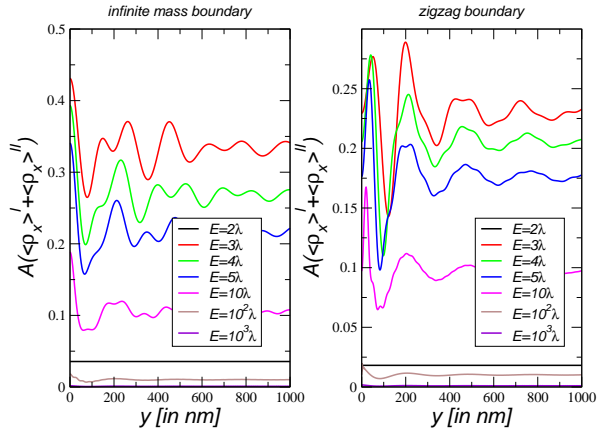}
\caption{Net spin polarization in $x$-direction as function of $y$ for various energies $E\geq2\lambda$. We use $\hbar v_F=5.6$eV\AA${}$ and $\lambda=6$meV. Left: Infinite mass boundary. Right: Zigzag boundary.}
  \label{fig:SpinDensity}
\end{center}
\end{figure}

In the following, we will study the spin polarization averaged over the incident direction for fixed $A_1,A_2$ and including the two $K$-points as function of the $y$-direction. We will further average over positive and negative $k_x$-momenta. With an incident wave of type $\mu$ and momentum $k^\mu=\sqrt{(E+(-1)^\mu|\lambda|)^2-\lambda^2}/(\hbar v_F)$, $\mu\in\{1,2\}$, we have 
\begin{equation} 
\langle\vec \rho\rangle^\mu(\vec r)\equiv\frac{1}{2}\sum_{\kappa=K,K'}\frac{1}{\pi}\int_{0}^\pi d\varphi \langle \psi_{E,k^\mu\cos\varphi}|\vec \sigma\delta(\vec r-\hat{\vec r})|\psi_{E,k^\mu\cos\varphi}\rangle_\kappa\;.
\end{equation}
We only discuss the spin polarization at the lower boundary which depends on the sign of $\lambda$ (here we choose $\lambda=|\lambda|$). The spin polarization on the upper boundary is obtained by reversing the sign.  

In Fig. \ref{fig:YDependenceAve}, the angle-averaged spin density $\mathcal{A}\langle\rho_x\rangle^\mu(\vec r)$ is shown as function of $y$ for various energies $E\geq2\lambda$ where $\mathcal{A}$ denotes the area of the sample. We show the results for an incident plane wave of type I (left hand side) and type II (right hand side) with an infinite mass boundary. There is a clear difference between the two types for low energies which is due to the appearance of imaginary momenta $k_y^1=\pm iq$ for type II-reflections. For low energies, most incident angles of the initial plane wave of type II lead to evanescent modes and thus to $\langle\sigma_x\rangle=0$. For large energies $E\geq10^3\lambda$, the spin polarization of the two types have approximately the same absolute value, but differ in sign.

Obviously, the above ensemble average breaks time-reversal symmetry since there is one incident plane wave with fixed $k_y$-direction and two reflected plane waves. But if there is no coherence between the incident plane waves of type I and II, e.g., due to temperature, then time-reversal symmetry is effectively broken and we find a net polarization in $x$-direction by adding the two contributions $\langle\vec \rho\rangle^I$ and $\langle\vec \rho\rangle^{II}$ (and possibly weighting them with the corresponding density of states). This is demonstrated in Fig. \ref{fig:SpinDensity}, where the sum of the two contributions ${\mathcal A}\sum_\mu\langle\vec \rho\rangle^{\mu}$ is shown for a infinite mass boundary (left) and for a zigzag boundary (right). Moreover, we expect spin polarization in $x$-direction for various non-equilibrium situations.

In the other two directions, we find no net spin polarization if the two inequivalent $K$-points are included. We note, however, that $\rho_y$ and $\rho_z$ assume a finite value for one $K$-point, only. This opens up the possibility of spin polarization in these directions in the presence of ripples or a magnetic field. Especially surface states due to, e.g., zigzag boundaries which effectively break the sublattice symmetry and which are not included in our continuous model should give rise to a finite spin polarization.

\section{Dirac electrons with ``pseudo-Rashba'' spin-orbit coupling in nanoribbons}
\label{sec:ribbon}
\begin{figure}[t]
\begin{center}
\includegraphics[angle=0,width=0.5\linewidth]{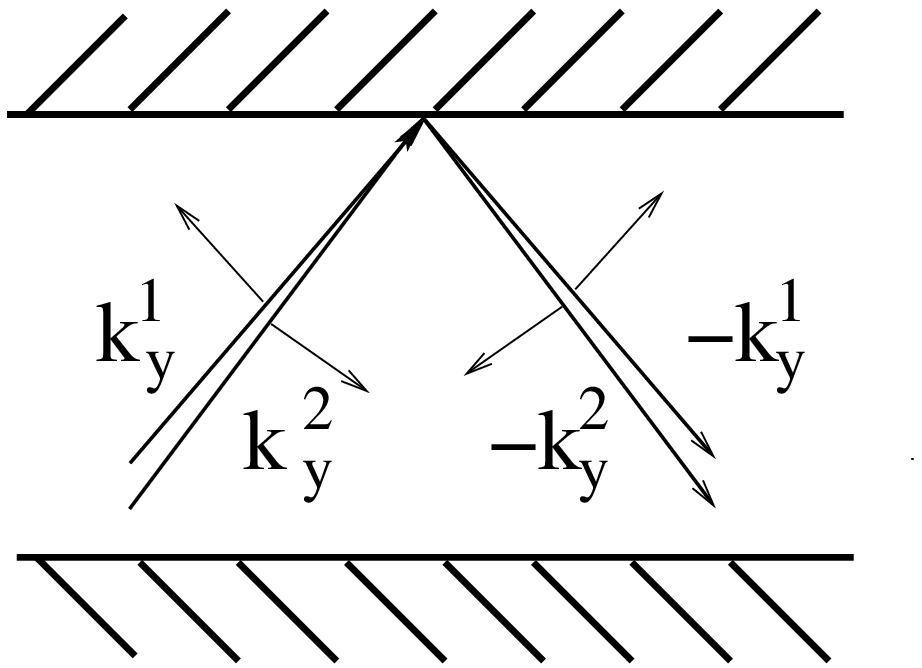}
\caption{A superposition of plane waves of type I and type II with constant $k_x$ reflected at one boundary of a nanoribbon into another superposition of pane waves of type I and type II.}
  \label{fig:ribbon}
\end{center}
\end{figure}

In this section, we will consider graphene nanoribbons and the quantization properties of the transverse momenta in the presence of ``pseudo-Rashba'' spin-orbit coupling. We will then discuss the density and spin distribution at various energies.

\subsection{Quantization of the transverse momentum}
Let us first consider infinite mass boundaries. For a general plane wave with fixed momentum $k_x$ and energy $E$, $\psi_{E,k_x}(x,y)\equiv(\psi_1,\psi_2,\psi_3,\psi_4)^T$, there are four conditions that have to be satisfied, i.e., $\psi_1=\pm\psi_3$ and $\psi_2=\pm\psi_4$ at $y=0$, and $\psi_1=\mp\psi_3$ and $\psi_2=\mp\psi_4$ at $y=W$, where the upper (lower) sign stands for the $K$($K'$)-point and $W$ the width of the nanoribbon. For a zigzag nanoribbon which terminates on sublattice $A$ at the bottom and on sublattice $B$ at the top, the four conditions read $\psi_1=\psi_2=0$ at $y=0$ and $\psi_3=\psi_4=0$ at $y=W$.

Let us first assume two propagating waves as in Eq. (\ref{wavefunction}), see also Fig. \ref{fig:ribbon}. In order to have a non-trivial solution, a necessary condition is
\begin{equation}
\label{BoundaryC}
\det M=
\det\left(
\begin{array}{cc}
 A & \bar A \\
 B & \bar B
\end{array}
\right)=\det\left(AB^{-1}-\bar A \bar B^{-1}\right)\det B\det\bar B=0\;,
\end{equation}
with the bar denoting the complex conjugate.

For infinite mass boundaries, the above matrices read at the $K$-point
\begin{equation}
A=
\left(
\begin{array}{cc}
 s_1 -c_1z_1& c_2-s_2z_2 \\
(c_1 -s_1z_1)z_1& (-s_2+c_2z_2)z_2
\end{array}
\right)\;,\;
B=
\left(
\begin{array}{cc}
(s_1+c_1z_1)w_1 & (c_2+s_2z_2)w_2\\
(c_1+s_1z_1)z_1w_1 & -(s_2+c_2z_2)z_2w_2 
\end{array}
\right)\;,
\end{equation}
and for zigzag boundaries, we have
\begin{equation}
A=
\left(
\begin{array}{cc}
 s_1 & c_2 \\
 c_1z_1 & -s_2z_2
\end{array}
\right)\;,\;
B=
\left(
\begin{array}{cc}
 c_1z_1w_1 & s_2z_2w_2\\
s_1z_1^2w_1 & -c_2z_2^2w_2 
\end{array}
\right)\;,
\end{equation}
where we introduced $w_\mu=e^{ik_y^\mu W}$ and used the definitions of section \ref{dephasing}. $\det M$ in Eq. (\ref{BoundaryC}) is real and thus yields the quantization of the transverse momentum in $y$-direction.

For $(\hbar v_{F}k_y^2)^2<4E\lambda$, there is the appearance of evanescent modes since $k_y^1=\pm iq$ is imaginary. In this case, a general plane wave with fixed momentum $k_x$ and energy $E\geq2|\lambda|$, $\psi_{E,k_x}(x,y)\equiv(\psi_1,\psi_2,\psi_3,\psi_4)^T$, is written as
\begin{eqnarray}
  \psi_{E,k_x}(x,y)&=&\mathcal{N}_{\vec k}e^{ik_xx}\Big[A_1e^{-q(W-y)}|\chi_{1,+}(k_x,-iq)\rangle +A_2e^{ik_y^2y}|\chi_{2,+}(k_x,k_y^2)\rangle\nonumber\\
    \label{wavefunctionEvanescentRibbon}
    &+&R_1e^{-qy}|\chi_{1,+}(k_x,iq)\rangle +R_2e^{-ik_y^2y}|\chi_{2,+}(k_x,-k_y^2)\rangle\Big]\;,
\end{eqnarray}
with
\begin{equation}
\label{Def:kyEvanescent}
\hbar v_Fq=\sqrt{-(E-|\lambda|)^2+\lambda^2+(\hbar vk_x)^2}\;,\;
\hbar v_Fk_y^2=\sqrt{(E+|\lambda|)^2-\lambda^2-(\hbar vk_x)^2}\;.
\end{equation}

Again, in order to have a non-trivial solution, Eq. (\ref{BoundaryC})
must hold, but this time the matrices for infinite mass boundaries at
the $K$-point read
\begin{equation}
A=
\left(
\begin{array}{cc}
(s_1 -c_1z_1^+)w_1& c_2-s_2z_2 \\
(c_1 -s_1z_1^+)z_1^+w_1& (-s_2+c_2z_2)z_2
\end{array}
\right)\;,\;
B=
\left(
\begin{array}{cc}
s_1+c_1z_1^+ & (c_2+s_2z_2)w_2\\
(c_1+s_1z_1^+)z_1^+ & -(s_2+c_2z_2)z_2w_2 
\end{array}
\right)\;,
\end{equation}
and for zigzag boundaries, we have
\begin{equation}
A=
\left(
\begin{array}{cc}
 s_1w_1 & c_2 \\
 c_1z_1^+w_1 & -s_2z_2
\end{array}
\right)\;,\;
B=
\left(
\begin{array}{cc}
 c_1z_1^+ & -s_2z_2w_2\\
s_1(z_1^+)^2 & -c_2z_2^2w_2 
\end{array}
\right)\;,
\end{equation}
with $w_1=e^{-qW}$, $z_1^\pm=(k_x\pm q)/\sqrt{k_x^2-q^2}$, $c_{1}\rightarrow\sqrt{(1+\cos\vartheta_1)/2}$ and $s_{1}\rightarrow i\sqrt{(\cos\vartheta_1-1)/2}$. The definitions for the plane wave of type II remain unchanged. 
Since the wave function of the evanecent mode is now real, the matrices
$\bar A$, $\bar B$ are not the complex conjugates of $A$, $B$, but given by
\begin{equation}
\bar A=
\left(
\begin{array}{cc}
s_1-c_1z_1^-& c_2-s_2z_2^* \\
(c_1-s_1z_1^-)z_1^-& (-s_2+c_2z_2^*)z_2^*
\end{array}
\right)\;,\;
\bar B=
\left(
\begin{array}{cc}
(s_1+c_1z_1^-)w_1 & (c_2+s_2z_2^*)w_2^*\\
(c_1+s_1z_1^-)z_1^-w_1 & -(s_2+c_2z_2^*)(z_2w_2)^* 
\end{array}
\right)\;,
\end{equation}
for infinite mass boundaries, and for zigzag boundaries they read
\begin{equation}
\bar A=
\left(
\begin{array}{cc}
  s_1 & c_2 \\
  c_1z_1^- & -s_2z_2^*
\end{array}
\right)\;,\;
\bar B=
\left(
\begin{array}{cc}
 c_1z_1^-w_1 & -s_2(z_2w_2)^*\\
 s_1(z_1^-)^2w_1 & -c_2(z_2^2w_2)^* 
\end{array}
\right)\;.
\end{equation}
It is now preferable to write Eq. (\ref{BoundaryC}) in powers of $w_1$. For zigzag boundaries, this yields 
\begin{eqnarray}
  &&\det M=2i{\rm Im}\left(\det B \det\bar A\right)+2w_1(z_1^+-z_1^-)(z_2-\bar z_2)s_1c_1s_2c_2\\\nonumber
  &-&w_1^2\Big[z_1^-(w_2z_2-\bar w_2\bar z_2)s_1c_1s_2c_2+(z_1^-)^2(w_2-\bar w_2)s_1^2s_2^2+(w_2z_2-\bar w_2\bar z_2)c_1^2c_2^2\Big]=0\;,
\end{eqnarray}
which is purely imaginary and thus again yields a quantization of the transverse momentum in $y$-direction. For infinite mass boundaries, we obtain a similar expression.

\subsection{Spin and density distribution}
\begin{figure}[t]
\begin{center}
\includegraphics[angle=0,width=0.5\linewidth]{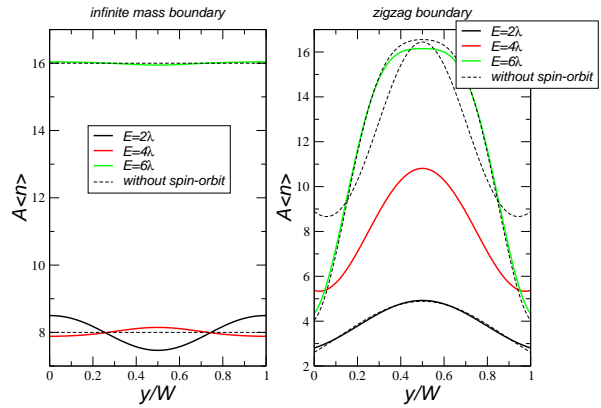}
\caption{The density distribution of a graphene nanoribbon of width $W=100$nm for various low energies. We use $\hbar v_F=5.6$eV\AA${}$ and $\lambda=6$meV. Left hand side: Infinite mass boundaries. Right hand side: Zigzag boundaries. The corresponding density distributions without ``pseudo-Rashba'' spin-orbit coupling are also shown (dashed lines).}
  \label{fig:DensityRibbon}
\end{center}
\end{figure}

The particle density $\langle n\rangle$ at energy $E$ is now obtained by summing over all transverse modes $n$ that obey the above boundary conditions or the corresponding boundary conditions for the $K'$-point. Denoting the $n$-th transverse momentum of type II  by $k_{y,n}^2$, we have
\begin{equation}
\langle n\rangle(\vec r)=\sum_{n}\sum_{k_x}\langle\psi_{E,k_x}|\delta(\vec r-\hat{\vec r})|\psi_{E,k_x}\rangle\delta_{E,E_{k_x,k_{y,n}^2}}\;.
\end{equation}

In Fig. \ref{fig:DensityRibbon}, the density distribution of a graphene nanoribbon of width $W=100$nm for various low energies with infinite mass (left) and zigzag (right) boundaries is shown. In general, the number of modes is the same with and without ``pseudo-Rashba'' spin-orbit coupling and the resulting density distributions only differ slightly. But for zigzag boundaries at $E=4\lambda$, we observe strong deviations due to the fact that there are 8 modes in the case 
with spin-orbit coupling in contrast to 12 modes in the case without spin-orbit coupling. Also note that whereas for the case without spin-orbit coupling, all modes are extended, some modes for the case with spin-orbit coupling are evanescent for the type I-branch. For zigzag boundaries, e.g., we have no extended and 4 evanescent (type-I) modes at $E=2\lambda$, 6 extended and 2 evanescent modes at $E=4\lambda$ and 6 extended and 6 evanescent modes at $E=6\lambda$. 

The spin polarization at the boundaries is in all cases zero; in $x$-direction it is zero also for only one $K$-point, in $y$- and $z$-direction it is non-zero for one $K$-point, but averages to zero when two $K$-points are included. This is an immediate consequence of time-reversal symmetry. In Ref. \cite{Zarea09}, spin polarization in $z$-direction is reported for various $k_x$-values within a lattice model of a zigzag nanoribbon. At equilibrium, this can only be attributed to edge-states which effectively break the sublattice symmetry and which are not included in our continuous model.

\section{Summary}
\label{summary}
In this paper, we have investigated the spin dephasing of Dirac fermions with ``pseudo-Rashba'' spin-orbit coupling due to the reflection from a hard wall. In order to confine the Dirac electrons, we used infinite mass and zigzag boundaries. For large energies compared to the spin-orbit coupling, we obtained the expected result that there is hardly spin-dephasing due to the scattering process. But for energies close to the band gap for plane waves of type I, $E\approx2\lambda$, strong spin dephasing is observed. If the incident plane wave is of type II (gapless branch), even stronger effects are seen like the appearance of evanescent modes. We also observe the rotation of the spin in out-of-plane direction away from the boundary and for incident plane waves of type II also at the boundary. This effect will be canceled by averaging over the two inequivalent $K$-points.

We also discussed the spin polarization averaged over the incident direction and including the two $K$-points. We find that for energies $E\geq2\lambda$, there is a finite spin polarization in $x$-direction when there is no coherence between the two branches. This polarization differs in sign for the upper and lower boundary, respectively. Also for non-equilibrium situations, there will be a spin polarization in this direction. 

We finally analyzed the spin and density distribution of graphene nanoribbons. At certain energies, the number of transverse modes does not match the one of a corresponding nanoribbon without ``pseudo-Rashba'' spin-orbit coupling. This results in significant changes in the density distribution. But generally, the ``pseudo-Rashba'' spin-orbit coupling leads to marginal differences, only. Further, there is no spin polarization if both K-points are included, but we find a finite spin polarization in $y$- and $z$-direction for one K-point, only. Surface states due to, e.g., zigzag boundaries which only live on one sublattice and thus break the valley-symmetry should therefore yield a finite spin polarization.  

\section{Acknowledgments}
T.S. wants to thank Nuno Peres and Jo\~ao Lopes dos Santos for illuminating discussions and support. We further thank E. Rashba for useful comments on the manuscript. This work was funded by FCT via the projects PTDC/FIS/64404/2006, PTDC/FIS/101434/2008 and by Deutsche Forschungsgemeinschaft via SFB 689. 

\section{Appendix A: The full model including the two $K$-points}
The full model including the two $K$-points reads
\begin{equation}
H=v\left(p_x\kappa_z\tau_x+p_y\tau_y\right)+\lambda\left(\kappa_z\tau_x\sigma_y-\tau_y\sigma_x\right)\;,
\end{equation}
where $\kappa_z=\pm1$ denotes the two inequivalent $K$-points. For a
given wave vector $\vec k$ the Hamiltonian around the $K'$-point
($\kappa_z=-1$ ) reads
\begin{equation}
{\cal H}(\vec k)=-
\left(
\begin{array}{cccc}
 0 & 0 & \hbar v_F(k_{x}+ik_{y}) & -2i\lambda \\
 0 & 0 & 0& \hbar v_F(k_{x}+ik_{y}) \\
\hbar v_F(k_{x}-ik_{y}) & 0 & 0 & 0\\
 2i\lambda & \hbar v_F(k_{x}-ik_{y}) & 0 & 0 
\end{array}
\right)\;.
\end{equation}

The Hamiltonian around the $K'$-point can thus be obtained from the Hamiltonian around the $K$-point by interchanging the pseudo-spin index and reversing the sign. All previous results without the mass term can thus be used. The results involving the mass term are obtained by $M\rightarrow-M$. This leads to a change in the boundary conditions, i.e.,
\begin{equation}
\label{BoundaryCondKprime}
\frac{\psi_1}{\psi_3}\Bigg|_{\rm bottom}^{K'}=\frac{\psi_2}{\psi_4}\Bigg|_{\rm bottom}^{K'}=-1\quad,\quad 
\frac{\psi_1}{\psi_3}\Bigg|_{\rm top}^{K'}=\frac{\psi_2}{\psi_4}\Bigg|_{\rm top}^{K'}=1\quad.
\end{equation}

\section{Appendix B: Massive Dirac fermions with ``pseudo-Rashba'' spin-orbit coupling}
Massive Dirac fermions with ``pseudo-Rashba'' spin-orbit interaction can be described by
\begin{equation}
{\cal H}=v\vec p\cdot\vec\tau+\lambda\left(\vec\tau\times\vec\sigma\right)
\cdot\vec e_{z}+Mv^2\tau_z\,,
\end{equation}
where, among standard notation, $\lambda$ is the spin-orbit coupling parameter,
and the Pauli matrices $\vec\tau$, $\vec\sigma$ describe the 
sublattice and the electron spin degree of freedom, respectively.

Squaring the Hamiltonian, we obtain the same eigenvectors as for massless Dirac fermions given in Eqs. (\ref{alpha_beta_1}) and (\ref{alpha_beta_2}). In the basis $(|\alpha_{1}\rangle,|\beta_{1}\rangle,|\alpha_{2}\rangle,|\beta_{2}\rangle)$ the Hamiltonian reads
\begin{equation}
\tilde{\cal H}(\vec k)=
\left(
\begin{array}{cccc}
 m & q_{+}^{\ast} & 0 & 0 \\
 q_{+} & -m  & 0 & 0 \\
 0 & 0 & m &  q_{-} \\
 0 & 0 & q_{-}^{\ast} & -m 
\end{array}
\right)
\end{equation}
with
\begin{equation}
q_{\pm}=\pm\hbar v_F(k_{x}\pm ik_{y})f_{\pm}(|\lambda|/ \hbar v_Fk)\qquad,\qquad m=Mv^2
\end{equation}
and 
\begin{equation}
f_{\pm}(x)=\sqrt{1+x^{2}}\pm x\,.
\end{equation}

Again we find two types of solutions. The first type has eigenvalues
\begin{equation}
\varepsilon_{1,\pm}=\pm\sqrt{M^2v^4+(\hbar v_Fk)^{2}+2\lambda^{2}
+2|\lambda|\sqrt{(\hbar v_Fk)^{2}+\lambda^{2}}}
\end{equation}
with eigenspinors
\begin{equation}
|\chi_{1,+}(\vec k)\rangle=
\left(
\begin{array}{c}
\sin(\vartheta/2)\cos(\zeta_1/2)\\
\cos(\vartheta/2)\cos(\zeta_1/2)e^{i\eta}\\
\cos(\vartheta/2)\sin(\zeta_1/2)e^{i\psi}\\
\sin(\vartheta/2)\sin(\zeta_1/2)e^{i\eta}e^{i\psi}
\end{array}
\right)\;,\;
|\chi_{1,-}(\vec k)\rangle=
\left(
\begin{array}{c}
\sin(\vartheta/2)\sin(\zeta_1/2)\\
\cos(\vartheta/2)\sin(\zeta_1/2)e^{i\eta}\\
-\cos(\vartheta/2)\cos(\zeta_1/2)e^{i\psi}\\
-\sin(\vartheta/2)\cos(\zeta_1/2)e^{i\eta}e^{i\psi}
\end{array}
\right)
\end{equation}
with $\zeta_{1/2}\in[0,\pi]$ and 
\begin{equation}
\cos\zeta_{1/2}=\frac{Mv^2}{\sqrt{|q_\pm|^2+M^2v^4}}\qquad,\qquad
e^{i\psi}=\frac{k_{x}+ik_{y}}{k}\,.
\end{equation}
The second type has eigenvalues
\begin{equation}
\varepsilon_{2,\pm}=\pm\sqrt{M^2v^4+(\hbar v_Fk)^{2}+2\lambda^{2}
-2|\lambda|\sqrt{(\hbar v_Fk)^{2}+\lambda^{2}}}
\end{equation}
with eigenspinors
\begin{equation}
|\chi_{2,+}(\vec k)\rangle=
\left(
\begin{array}{c}
\cos(\vartheta/2)\cos(\zeta_2/2)\\
-\sin(\vartheta/2)\cos(\zeta_2/2)e^{i\eta}\\
\sin(\vartheta/2)\sin(\zeta_2/2)e^{i\psi}\\
-\cos(\vartheta/2)\sin(\zeta_2/2)e^{i\eta}e^{i\psi}
\end{array}
\right)\;,\;
|\chi_{2,-}(\vec k)\rangle=
\left(
\begin{array}{c}
\cos(\vartheta/2)\sin(\zeta_2/2)\\
-\sin(\vartheta/2)\sin(\zeta_2/2)e^{i\eta}\\
-\sin(\vartheta/2)\cos(\zeta_2/2)e^{i\psi}\\
\cos(\vartheta/2)\cos(\zeta_2/2)e^{i\eta}e^{i\psi}
\end{array}
\right)
\,.
\end{equation}
Let us now consider expectation values within the eigenstates with
wave functions
\begin{equation}
\langle\vec r|\vec k,\mu,\pm\rangle=\frac{e^{i\vec k\vec r}}{\sqrt{\cal A}}
|\chi_{\mu,\pm}\rangle\,,
\end{equation}
$\mu\in\{1,2\}$, and $\cal A$ being the area of the system.
Here we find
\begin{equation}
\langle\vec k,\mu,\pm|\vec\tau|\vec k,\mu,\pm\rangle
=\pm\left(
\begin{array}{c}
\sin\vartheta\sin\zeta_{\mu}\cos\varphi\\
\sin\vartheta\sin\zeta_{\mu}\sin\varphi\\
\cos\zeta_{\mu}
\end{array}
\right)\,,
\end{equation}
and
\begin{equation}
\langle\vec k,1,\pm|\vec\sigma|\vec k,1,\pm\rangle
=-\langle\vec k,2,\pm|\vec\sigma|\vec k,2,\pm\rangle
=\left(
\begin{array}{c}
-\sin\vartheta\sin\varphi\\
\sin\vartheta\cos\varphi\\
\mp\cos\vartheta\cos\zeta_{1/2}
\end{array}
\right)\,.
\end{equation}
Here we have assumed a positive spin-orbit coupling parameter, 
$\lambda=|\lambda|$, and $\varphi$ is the usual azimuthal angle of the wave 
vector, $\vec k=k(\cos\varphi,\sin\varphi)$. Note that massive Dirac fermions assume a non-zero expectation value for the pseudo-spin and spin in $z$-direction. 

\section{Appendix C: Scattering from infinite mass boundary}
Dirac fermions can be confined by an infinite mass boundary, first discussed by Berry and Mondragon \cite{Berry87}. In the following, we will study the scattering behavior from a boundary located at $y=0$ and $y=W$. Within the strip $0<y<W$, the mass of the Dirac fermions shall be zero; outside the strip, the mass shall be infinite. 

A general plane wave within the strip with fixed momentum $k_x$ and energy $E>0$ can be written as 
\begin{eqnarray}
  \psi_{E,k_x}(x,y)&=&e^{ik_xx}\Big[A_1e^{ik_y^1y}|\chi_{1,+}(k_x,k_y^1)\rangle +A_2e^{ik_y^2y}|\chi_{2,+}(k_x,k_y^2)\rangle\nonumber\\
\label{app:psi}
    &+&R_1e^{-ik_y^1y}|\chi_{1,+}(k_x,-k_y^1)\rangle +R_2e^{-ik_y^2y}|\chi_{2,+}(k_x,-k_y^2)\rangle\Big]\;,
\end{eqnarray}
with
\begin{equation}
\hbar v_Fk_y^{\mu}=\sqrt{(E+(-1)^\mu|\lambda|)^2-\lambda^2-(\hbar v_Fk_x)^2}\;,
\end{equation}
$\mu\in\{1,2\}$.

The wave function of the transmitted electron is also decomposed by the two eigenfunctions $|\chi_{\mu,+}\rangle$,
\begin{equation}
\widetilde\psi_{E,k_x}(x,y)=e^{ik_xx}\Big[T_1e^{ik_y^1y}|\chi_{1,+}(k_x,k_y^1)\rangle +T_2e^{ik_y^2y}|\chi_{2,+}(k_x,k_y^2)\rangle\Big]\;,
\end{equation}
with
\begin{equation}
\hbar v_Fk_y^{\mu}=\sqrt{(\sqrt{E^2-M^2v^4}+(-1)^\mu|\lambda|)^2-\lambda^2-(\hbar v_Fk_x)^2}\;.
\end{equation}

In the limit $M\rightarrow\infty$, the transmitted plane wave simplifies 
\begin{equation}
\widetilde\psi_{E,k_x}(x,0)
=e^{ik_xx}\widetilde T_1\left(
\begin{array}{c}
1\\
-s_\lambda\\
1\\
-s_\lambda
\end{array}
\right)
+e^{ik_xx}\widetilde T_2\left(
\begin{array}{c}
1\\
s_\lambda\\
1\\
s_\lambda
\end{array}
\right)
\end{equation}
and 
\begin{equation}
\widetilde\psi_{E,k_x}(x,W)
=e^{ik_xx}\widetilde T_1\left(
\begin{array}{c}
1\\
-s_\lambda\\
-1\\
s_\lambda
\end{array}
\right)
+e^{ik_xx}\widetilde T_2\left(
\begin{array}{c}
1\\
s_\lambda\\
-1\\
-s_\lambda
\end{array}
\right)\;,
\end{equation}
with $s_\lambda=\lambda/|\lambda|$. The different expressions at $y=0$ and $y=W$ originate from the different sign of $\hbar v_Fk_y\rightarrow\pm iMv^2$ that has to be chosen to yield an exponential decay in the infinite mass region. It therefor only depends on whether one deals with the upper or lower boundary.

At the boundaries $y=0$ and $y=W$, the four components have to be continuous to guarantee a continuous current
which leads to the following two sets of equations:
\begin{equation}
\label{MatchingConditions}
\psi_{E,k_x}(x,0)=\widetilde\psi_{E,k_x}(x,0)\qquad,\qquad\psi_{E,k_x}(x,W)=\widetilde\psi_{E,k_x}(x,W)
\end{equation}

With $\psi_{E,k_x}=(\psi_1,\psi_2,\psi_3,\psi_4)^T$, this translates to the familiar boundary condition from Ref. \cite{Berry87} for the two spin channels, respectively:
\begin{equation}
\frac{\psi_1}{\psi_3}\Bigg|_{y=0}=\frac{\psi_2}{\psi_4}\Bigg|_{y=0}=1\quad,\quad 
\frac{\psi_1}{\psi_3}\Bigg|_{y=W}=\frac{\psi_2}{\psi_4}\Bigg|_{y=W}=-1\quad.
\end{equation}

%To obtain the reflection coefficients $R_1$ and $R_2$, it is convenient to multiply the reflected waves in Eq. (\ref{app:psi}) with $z_{\mu}^2$ where $z_\mu=(k_x+ik_y^{\mu})/\sqrt{k_x^2+(k_y^{\mu})^2}$, $\mu\in\{1,2\}$.

\end{document}